\documentclass[useAMS,usenatbib]{mn2e}
\usepackage{amssymb}
\usepackage{epsfig}
\usepackage{Times}

\title[Extremely compact  massive galaxies at z$\sim$1.4]
{Extremely compact  massive galaxies at z$\sim$1.4}

\author[I. Trujillo et al.]{I. Trujillo$^{1}$, G. Feulner$^{2,3}$\footnote{Current address:
  Potsdam--Institut f\"ur Klimafolgenforschung, Postfach 60~12~03,
  D--14412 Potsdam, Germany},
 Y. Goranova$^{2,3}$, U.
Hopp$^{2,3}$,  M. Longhetti$^{4}$, P.  Saracco$^{4}$,\newauthor
 R. Bender$^{2,3}$, V.
Braito$^{4}$,  R. Della Ceca$^{4}$, N. Drory$^{5}$, 
F. Mannucci$^{6}$ and P.
Severgnini$^{4}$\\
 $^{1}$School of Physics and Astronomy, University of Nottingham,
   University Park, Nottingham NG7 2RD, UK\\
$^{2}$Universit\"ats-Sternwarte M\"unchen, Scheinerstrasse 1, D-81679
	     M\"unchen, Germany\\
$^{3}$
 Max-Planck-Institut f\"ur extraterrestriche Physik, Giessenbachstrasse,
 D-85748 Garching, Germany\\
$^{4}$INAF - Osservatorio Astronomico de Brera, Via Brera 28, 20121 Milano,
	 Italy\\
	 $^{5}$
 University of Texas at Austin, Austin, Texas 78712, USA\\
 $^{6}$
 IRA-CNR, Firenze, Italy}

\begin{document}

\date{}

\pagerange{\pageref{firstpage}--\pageref{lastpage}} \pubyear{2002}

\maketitle

\label{firstpage}

\vspace{-1cm}

\begin{abstract}

The optical rest--frame sizes of 10 of the most massive
($\sim$5$\times$10$^{11}$h$_{70}$$^{-2}$M$_{\sun}$)  galaxies found in the
near--infrared MUNICS survey at 1.2$<$z$<$1.7 are analysed. Sizes were estimated
both in the J and K' filters. These massive galaxies are at least a factor of 
4$_{-1.0}^{+1.9}$ ($\pm$1 $\sigma$) smaller in the rest--frame V--band than local
counterparts of the same stellar mass. Consequently, the stellar mass density of
these objects is (at least) 60 times larger than massive ellipticals today.
Although the stellar populations of these objects are passively fading, their
structural properties are rapidly changing since that redshift. This observational
fact disagrees with a scenario where the more massive and passive galaxies are
fully assembled at z$\sim$1.4 (i.e. a monolithic scenario) and  points
towards a dry merger scenario as the responsible mechanism for the subsequent
evolution of these galaxies.  

\end{abstract}


\vspace{-1cm}
\section{Introduction}

In the local universe, the population of galaxies with stellar masses larger
than 10$^{11}$M$_{\sun}$ is dominated by passive early--type galaxies that are
a factor of $\sim$3 more numerous than late--type galaxies above this mass
threshold (Baldry et al 2004).  Their stellar populations are old, metal rich
and are characterized by a short formation time-scale (e.g. Heavens et al.
2004; Thomas et al 2005; Feulner et al. 2005). In addition, these massive
galaxies are large, with sizes (as parametrized by the effective radius) larger
than 4 kpc (Shen et al. 2003). 

Historically, two different formation scenarios have been proposed in order to
explain the properties of these objects: the so--called monolithic collapse model
(Eggen, Lynden--Bell \& Sandage 1962; Larson 1975; Arimoto \& Yoshii 1987; Bressan,
Chiosi \& Fagotto 1994) and the hierarchical merging model (White \& Frenk 1991).
In the former scenario, spheroidal galaxies formed at a very early epoch as a
result of a global starburst, and then passively evolve to the present. In the
merger model, spheroids are formed by violent relaxation during major merger
events.

Favoring the monolithic model is the fact that the bulk of stars in massive
ellipticals are old (Mannuci et al. 2001) and have high [$\alpha$/Fe] ratios (i.e.
short star formation time scales; Worthey, Faber \& Gonzalez 1992). On the other
hand, supporting a hierarchical merger scenario, current observations seem to find
a decline in the number of massive galaxies at high--z. The space density of red
passively evolving early--type galaxies has moderately increased since z$\sim$1
(Daddi et al. 2000; Pozzetti et al. 2003; Bell et al. 2004; Drory et al. 2004;
Faber et al. 2005). At even higher redshift, z$\sim$1.7, their space density
appears to be a factor of 2--3 smaller than that of their local counterparts
(Daddi et al. 2005; Saracco et al. 2005; Drory et al. 2005). In addition, new
generations of semianalytical models (e.g. de Lucia 2006) are now able to produce
results in better agreement with the stellar population properties of 
ellipticals.

An additional test to check the validity of the above two scenarios is to
explore the size evolution of the spheroids through  cosmic time. In this
sense, the prediction from the monolithic model is that the stellar mass--size
relation of these objects should remain unchanged after their formation, with
the luminosity--size relation evolving in agreement with the fading of their
stellar populations. In the hierarchical model, however, the stellar mass--size
relation of these objects should change as a result of the increase in size of
the remnants after each merger. 

From the observational point of view,  the evolution of the luminosity--size and
stellar mass--size relations of the early--type galaxies since z$\sim$1 is
consistent with the passive aging of ancient stellar populations (Trujillo \&
Aguerri 2004, McIntosh et al. 2005). Consequently, the structural properties of
the more massive galaxies   seem  to not change from z$\sim$1 to the present. At
z$>$1 the situation is less clear. Large and deep near--infrared surveys are
needed to collect a significative sample of massive passively evolving  galaxies
at z$>$1 and to explore the sizes of the galaxies in their optical rest--frame.
Using the Faint Infrared Extragalactic Survey (FIRES; Franx et al. 2000), Trujillo
et al. (2004;2006) have shown that there is a hint for most massive galaxies
(M$_{\star}$$\gtrsim$7$\times$10$^{10}$ h$_{70}$$^{-2}$M$_\odot$) at z$\sim$2.5
being a factor of 2 smaller than present--day counterparts with similar masses.
However, the number of galaxies in that sample is not large enough to firmly
establish this result. In addition, there has been a recent claim of 4 very
compact (r$_e$$\lesssim$1 kpc in their local UV restframe) and massive
(M$_{\star}$$>$10$^{11}$ h$_{70}$$^{-2}$M$_\odot$) passively evolving galaxies at 
z$\sim$1.7 in the UDF (Daddi et al. 2005). The few number of objects together with
the fact that some of these galaxies could be hosting an AGN make the above claims
uncertain.

To shed some light on the above issue, we estimate the sizes (in their local
optical rest--frame) of a sample of 10 very massive
(10$^{11}$$<$M$_{\star}$$<$10$^{12}$M$_{\sun}$)  galaxies spectroscopically
classified as early--type galaxies at 1.2$<$z$<$1.7, with no sign of AGN activity
in their spectra (except for one object). These objects are consequently good
candidates to test whether their sizes are equal or smaller than their local
counterparts.  Throughout, we will assume a flat $\Lambda$--dominated cosmology
($\Omega_M$=0.3, $\Omega_{\Lambda}$=0.7 and $H_0$=70  km s$^{-1}$ Mpc$^{-1}$). All
magnitudes are provided in the Vega system unless otherwise stated.

\vspace{-0.7cm}
\section[]{Description of the data}

The near-infrared images used in this study  were taken from the Munich Near-IR
Cluster Survey (MUNICS; Drory et al. 2001),  and its deeper follow-up project
called MUNICS-Deep. MUNICS is a wide-field medium-deep photometric survey taken in
the near-infrared and optical filters. Dedicated follow-up spectroscopy is
available for a selected sub-sample (Feulner et al. 2003). The main part of the
survey consists of 10 fields with a total area of $\sim 0.3$~deg$^2$. For all
these fields photometry in $K'$, $J$, $I$, $R$, $V$, and $B$ is available, with
limiting magnitudes ranging from K' $\simeq$ 19.5 and J $\simeq$ 21 to B $\simeq$
24.0 mag (50\% completeness for point sources; Snigula et al. 2002). This is
sufficiently deep to detect passively evolving systems up to a redshift of
z$\lesssim$1.4, and at a luminosity of 0.5L$^{*}$. The final $K'$-selected catalog
contains roughly 5000 objects and is described  in Drory et al. (2001).

The sample of massive Extremely Red Objects (ERO) studied here was selected in
three survey fields (S2F1, S2F5 and S7F5). The primary selection criteria were
$K'<18.5$ and $R-K'>5$, resulting in a list of 36 objects. Low-resolution
near-infrared spectroscopy was carried out for  $\sim 60$\% of them, and ten
objects are identified as early-type galaxies at  $1.2<z<1.7$, with
stellar masses well exceeding $10^{11} M_\odot$ (Saracco et al. 2003, 2005;
Longhetti et al. 2005).

Seven out of the ten EROs explored here are located in the field S2F1, for which
deeper near-infrared images are available from the MUNICS-Deep survey (Goranova
et al., in prep.). MUNICS-Deep aims at obtaining a contiguous 1-square-degree
field (overlapping the MUNICS patchs S2F1 and S2F5) in optical and near-infrared
filters to a detection limit 2~mag deeper than MUNICS. To improve the size
measurements of these seven EROs in the field S2F1, the deep $K'$ and $J$-band
images from MUNICS-Deep were used. These were obtained with Omega2000 at the
Calar Alto 3.5-m telescope at a pixel scale of 0.45 arcsec/pixel, a typical
seeing of $\lesssim$1.0 arcsec, and limiting magnitudes of K'$\sim$21.5 and
J$\sim$23.5 mag (again 50\% completeness limits for point sources). Basic data
reduction (Goranova et al., in prep.) was performed using a modified version of
the \texttt{IRAF} external package  \texttt{XDIMSUM}.

For the remaining three objects, the sizes were estimated on the $K'$-band and
J-band images of the (shallower) MUNICS project. These images, taken with
OmegaPrime at the Calar Alto 3.5-m telescope, have a pixel scale of 0.396
arcsec/pixel, a typical seeing of $\sim$1.2 arcsec, and a limiting magnitude of
K'$\sim$ 19.5. 

Stellar masses have been derived from the K'-band absolute magnitudes by means
of the mass--to--light ratio M/L$_{K'}$ derived from the best fitting models
(see details in Longhetti et al. 2005). The largest uncertainty in the stellar
mass computation comes from the variation of M/L according to the age of the
stellar population and the adopted IMF. However, it is worth noting that, given
the extremely bright K'-band  magnitudes (K'$<$18.4) and the redshifts (z$>$1.2)
of our galaxies, their resulting stellar masses are well in excess of 10$^{11}$
M$_{\sun}$ leaving aside any model assumption.  In this paper we use stellar
masses derived using a Kroupa IMF. However, to estimate the uncertainty in the
stellar masses we used a large set of different IMFs (Longhetti et al. 2005).

\vspace{-0.7cm}
\section[]{Size estimation}

To estimate the sizes of the galaxies we have used the GALFIT code (Peng et al.
2002). GALFIT convolves S\'ersic (1968) r$^{1/n}$ profile galaxy models with the point-spread
function (PSF) of the images and then determines the best fit by comparing the
convolved models with the science data using a Levenberg--Marquardt algorithm to
minimize the $\chi^2$ of the fit. 

The spatial resolution of our images does not allow to estimate accurately the
shape (index $n$) of the surface brightness profiles. For that reason, and to
decrease the number of free parameters in our fits, we have calculated the size
of the galaxies by fixing the S\'ersic index to n=1 (i.e. an exponential
profile) and n=4 (i.e. a de Vaucouleurs profile). Both models are convolved
with the image PSF. The effective radii provided by every fit (r$_{e,1}$ and
r$_{e,4}$) are used to estimate a mean effective radius and indicate the range
of variation of the sizes of our galaxies. The PSF that was used for every
galaxy corresponds to the nearest  (bright enough but non-saturated) star to
the galaxy.

Neighboring galaxies were excluded from each model fit using a mask, but in the
case of closely neighboring galaxies with overlapping isophotes, the galaxies
were fitted simultaneously.

\vspace{-0.35cm}
\subsection{Testing the size estimates: simulations}

The results presented in this paper rely on our ability to measure accurate
structural parameters. To gauge the accuracy of our size determination we have
created 250 artificial galaxies uniformly generated at random in the following
ranges: 18$\leq$ J$\leq$21, 0.$''$1$\leq$r$_e$$\leq$1.6$''$ (i.e.
0.8$\lesssim$r$_e$$\lesssim$13.5 h$_{70}^{-1}$ kpc in the local restframe at
z$\sim$1.4) and 0.5$\leq$n$\leq$8. Simulations were done in J band only, but our
results can be extrapolated to the K band data because of their similar
signal--to--noise quality. The mock galaxies span a large range of surface
brightness shapes (i.e. they are not restricted to n=1 or n=4) to model the
different galaxy profiles found in the observations (Trujillo et al. 2006). To
simulate the real conditions of our observations, we add a background sky image
(free of sources) taken from a piece of the MUNICS--Deep field  image in the J
band. Finally, the galaxy models used (n=1 and n=4) were convolved with the
observed PSF. The same procedure was used to retrieve the structural parameters
both in the simulated and actual images.

\begin{figure}
\epsfig{file=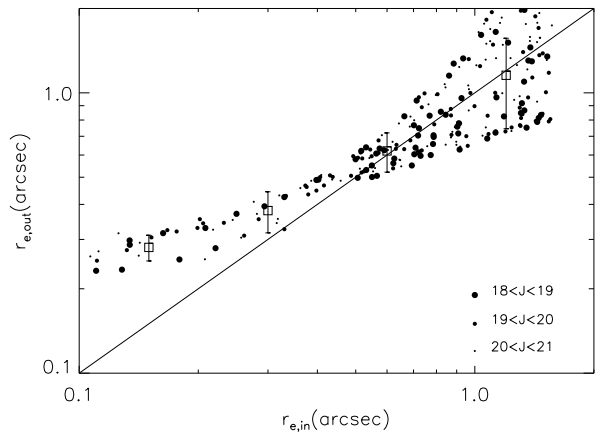,width=0.5\textwidth} 

\caption{The figure shows a comparison between the input intrinsic half--light
radius (before seeing convolution) and recovered size values in our simulations
for the MUNICS-Deep observations. The small points are used to indicate the
position of individual mock galaxies. The open squares indicate the mean value
and the bars correspond to 1$\sigma$ dispersion. According to the simulations,
the retrieved sizes for the smallest objects should be considered only as an
upper limit.}

\label{simulation}
\end{figure}

Fig. \ref{simulation} shows the comparison between the input and recovered size
values in our simulations. The recovered size value, r$_{e,out}$ is evaluated as
the mean value between the size recovered using n fixed to 1, and n fixed to 4:
r$_{e,out}$=(r$_{e,out,n=1}$+r$_{e,out,n=4}$)/2. For that reason, the intrinsic
scatter shown in Fig. \ref{simulation} is independent of the magnitudes and
 mainly caused by the fact that the index $n$ of the model galaxies is fixed
whereas the mock galaxies span a large range of n. According to the
simulations, our sizes for the smallest objects should be considered only as an
upper limit. This systematic deviation of the size of the galaxies at small
radii  is probably an artefact due to the relative large size of the pixel
($\sim$0.45\arcsec) compared to the size of the galaxies. Simulations also
provide us with a typical uncertainty in the estimation of the sizes of small
galaxies of $\sim$0.1\arcsec. We will use this value to estimate the error bars
in our measurements.

\vspace{-0.35cm}
\subsection{Testing the size estimates: crude upper limits and 
the effect of different images depth}

A direct method of establishing a crude upper limit to the size of the objects is
by fitting a Gaussian profile to the observed galaxies. Under the assumption that
the intrinsic surface brightness profile of the galaxies are also described by a
Gaussian profile, the effective radius of the galaxy is given by:

\begin{equation}
r_e=1/2\times\sqrt{FWHM_{obj}^2+FWHM_{PSF}^2}
\end{equation}

\noindent In general, galaxies have a surface brightness profile which is much
more concentrated than a Gaussian profile, consequently, this method  provides
us with an upper (very conservative) limit of the object's size. A direct
Gaussian fit to our objects provide a typical value of
FWHM$_{obj}$$\sim$1.2\arcsec. Using the typical value of seeing that we have in
our images, FWHM$_{PSF}$$\sim$1\arcsec, this translates into an upper limit to
the sizes of the galaxies of r$_e$$\lesssim$0.8\arcsec (or r$_e$$\lesssim$6.5
h$_{70}^{-1}$ kpc at z$\sim$1.4 in the cosmology used). Local galaxies with 
M$_{\star}$$\sim$5$\times$10$^{11}$h$_{70}^{-2}$M$_{\sun}$ are expected to have
sizes of $\sim$10 h$_{70}^{-1}$ kpc (Shen et al. 2003). This crude upper limit
estimation (according to our simulations this tecnique will produce estimates
$\sim$1.5 larger than the input values) shows that our high--z
massive galaxies are more compact than their local counterparts.

The sizes of three galaxies in our sample were estimated using shallower images
than  the rest of the sample. We have checked whether these shallow observations
could introduce any bias in the size estimates of these three objects. For
the seven galaxies where we have both shallow and deep observations we estimated
the sizes in both cases, and the sizes agree within the error bars. We do not
observe, in addition, any systematic difference. This result implies that using
deeper observations does not unveil the contribution of light from missing wings
in the surface brightness distributions. In other words, our images are catching
almost all the light of these galaxies.

\vspace{-0.9cm}
\section{The observed stellar mass vs size relation}

We have estimated the sizes of our galaxies in both the J and  the K' band. At
z$\sim$1.4, this implies estimating the sizes in the local rest--frame V--band and
(approximately) I--band. The results of our fitting are shown in Table \ref{data}.
The seeing and the depth are slightly different amongst the near infrared images
which allows us to test the reliability of the size estimates. Interestingly,
the sizes of individual objects both in J and in K' band are very similar. This
reinforces the idea that the sizes presented here are robust.

\begin{table*}
 \centering
 \begin{minipage}{140mm}
  \caption{Main physical parameters of the 10 early--type galaxies. Note:
  sizes of galaxies marked with an asterisk were obtained in the shallow MUNICS images}
  \begin{tabular}{rrrrrrrrrrrr}
  \hline
Field & ID & z$_{spec}$ & J  &  K'  & M$_{\star}$ & r$_{e,J,n=1}$ & 
r$_{e,J,n=4}$ & r$_{e,K,n=1}$ & r$_{e,K,n=4}$ & FWHM & FWHM \\
      &    &   &  &  &  & 
      &  &  &  & J-band & K'-band \\
      &    &   & (mag) & (mag) & (10$^{11}$M$_\odot$) & ($\arcsec$)
      & ($\arcsec$) & ($\arcsec$) & ($\arcsec$) & ($\arcsec$) & ($\arcsec$) \\
 \hline
S2F5* & 109 & 1.22 & 18.2 & 16.6 & 7.0 & 0.46 & 0.49 & 0.52 & 0.62 & 1.17 & 1.09 \\
S7F5* & 254 & 1.22 & 19.8 & 17.8 & 7.2 & 0.66 & 0.53 & 0.79 & 0.75 & 1.30 & 1.12 \\
S2F1  & 357 & 1.34 & 19.5 & 17.8 & 9.0 & 0.32 & 0.34 & 0.30 & 0.35 & 0.98 & 0.97 \\
S2F1  & 527 & 1.35 & 20.4 & 18.3 & 3.6 & 0.14 & 0.09 & 0.17 & 0.15 & 1.00 & 0.99 \\
S2F1  & 389 & 1.40 & 20.3 & 18.2 & 4.6 & 0.21 & 0.18 & 0.21 & 0.34 & 1.00 & 1.00 \\
S2F1  & 511 & 1.40 & 19.8 & 18.1 & 1.7 & 0.26 & 0.23 & 0.26 & 0.41 & 0.95 & 0.95 \\
S2F1  & 142 & 1.43 & 19.6 & 17.8 & 5.9 & 0.31 & 0.38 & 0.28 & 0.24 & 0.97 & 1.00 \\ 
S7F5* &  45 & 1.45 & 19.6 & 17.6 & 4.7 & 0.45 & 0.53 & 0.56 & 0.74 & 1.14 & 1.18 \\
S2F1  & 633 & 1.45 & 20.0 & 18.2 & 5.5 & 0.28 & 0.32 & 0.35 & 0.52 & 0.95 & 0.95 \\
S2F1  & 443 & 1.70 & 20.5 & 18.4 & 6.2 & 0.33 & 0.39 & 0.37 & 0.40 & 0.99 & 0.95 \\
\hline
\label{data}
\end{tabular}
\end{minipage}
\end{table*}

\begin{figure*}
\centering
\epsfig{file=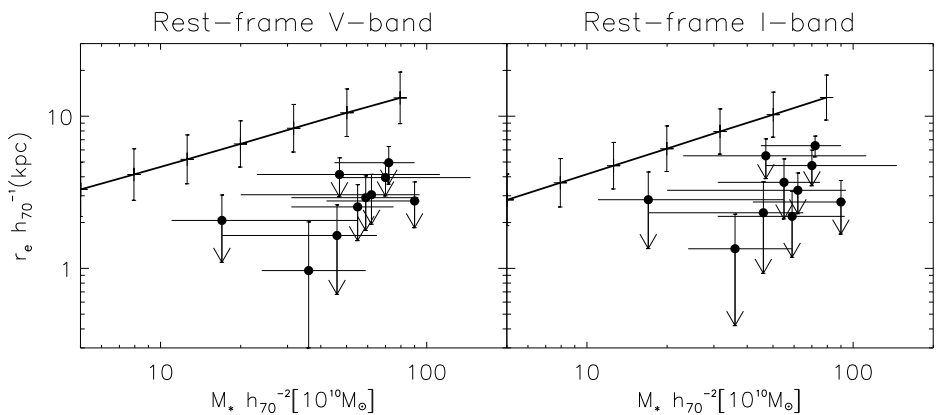,width=0.8\textwidth} 

\vspace{-4cm}

 \caption{Distribution of rest--frame optical sizes vs.  stellar mass  for massive
MUNICS galaxies. $Left$ $Panel$ shows the distribution of galaxies with sizes
estimated in the V-band local restframe. $Right$ $Panel$ shows the distribution of
galaxies with sizes estimated in the I-band local restframe. Overplotted on the
observed distribution of points are the mean and dispersion of the distribution of
the S\'ersic half--light radius of the SDSS galaxies as a function of the stellar
mass. SDSS sizes were obtained in the "V--band" and in the "I--band".}

\label{results}
\end{figure*}

The stellar mass--size relation for the massive galaxies analysed here are
presented in Fig. \ref{results}. Overplotted in this figure are the mean and
dispersion of the distribution of the S\'ersic half--light radii of early--type
galaxies from the Sloan Digital Sky Survey (SDSS; York et al. 2000). We use the
SDSS sample as the local reference. Local sizes are determined from a S\'ersic
model fit (Blanton et al. 2003) and the characteristics of the sample  described
in Shen et al. (2003). SDSS stellar masses were also derived using a Kroupa IMF.
The mean of the SDSS galaxies redshift distribution used in this comparison is
0.1. We use the sizes estimated in the observed r'--band and the z'--band (S.
Shen, private communication). This closely matchs the V--band and I--band
restframe filters at z$\sim$0.1.

Fig. \ref{results} shows that, at a given stellar mass, the most massive
galaxies at z$\sim$1.4 are much  smaller than local ones.  According to our
simulations our sizes are upper limits, impliying that our high--z galaxies are
at least a factor of 4.0$_{-1.0}^{+1.9}$ ($\pm$1 $\sigma$) smaller in the
V--band, and at least 3.2$_{-0.8}^{+1.8}$ ($\pm$1 $\sigma$) smaller in the
I--band than local counterparts. This implies that the internal stellar mass
density in the most massive galaxies at that redshift is $\gtrsim$60 (or at
least 33 if considered the measurements obtained in K'-band) times larger than
today. To test the robustness of our results we have checked two potential
biases. First, following Maraston et al. (2006), we repeat the analysis under
the assumption that our masses could be overestimated by a factor of $\sim$2. In
this case, our galaxies will still be more compact than present-day galaxies of the
same masses by a factor of 2.5-3. Second, following the fact that present very
massive ellipticals have large index $n$ values, we repeat our analysis using
the r$_e$ values obtained forcing the index  n to be fixed to 8 during the fitting. In this case,
our galaxies are still more compact than the local galaxies (of equal mass) by a
factor of 2.7-3.3.

Most of our galaxies are more than 2 $\sigma$ away from the local relation. In
fact, we have probed whether there is any galaxy in the SDSS sample as massive
and compact as the ones we have found. Using the catalogue used by Shen et al.
(2003) to build the local SDSS relations (S. Shen, private communication) we
have not found any local galaxy with r$_e$$<$4 kpc and
M$_{\star}$$>$3$\times$10$^{11}$M$_{\sun}$, and only one with r$_e$$<$5 kpc (see
also Bernardi et al. 2006). That means a comoving density of 10$^{-7}$Mpc$^{-3}$.
The selection of our objects (including the spectroscopic follow up) is not
biased towards smaller objects. In fact,  equally bright objects with sizes
four  times larger could be  observed, and their sizes measured accurately if
they were in our sample (see Fig. 1). When  considering all the galaxies
together the possibility that they are all  small by chance is  rejected at the
4 $\sigma$ level. 

\vspace{-0.9cm}
\section{Discussion}

As stated in the Introduction, it is possible to find in previous works some
examples of significantly small massive galaxies at z$\gtrsim$1. Daddi et al.
(2005) have been the first on discussing in detail the nature of these objects.
In addition to the four Daddi et al. objects, there is a group of compact
(r$_e$$\lesssim$1 kpc) and  massive galaxies at z$\sim$1 seen in Fig. 9 from
McIntosh et al. 2005. There is also  evidence of compact massive galaxies in
Trujillo et al. (2006) and di Serego Alighieri et al. (2005). Finally,
Waddington et al. (2002) studied two z$\sim$1.5 radio--selected early--type
galaxies and found, using NICMOS imaging, sizes of $\sim$0.3 arcsec ($\sim$ 2.5
kpc). Therefore, the observational existence of these small galaxies appears
currently to be well established.

A strong morphological K--correction has been suggested by Daddi et al. (2005)
as one of the potential explananation of the compacteness of their objects
observed in the UV restframe. However, our observations in the optical restframe
reject this possibility. Another potential explanation suggested by Daddi et al.
(2005) is the pressence of an unresolved nuclear component (i.e. an AGN). In
fact, two of their 4 objects were detected in X--rays. However, the AGN
hypothesis is unlikely to explain our observations. Only one of our galaxies
(S2F1 443) is detected in deep XMM-Newton pointing of the S2 fields (Severgnini
et al. 2005) having L$_{2-10keV}$$\gtrsim$10$^{43}$ergs s$^{-1}$ and the
spectral energy distributions of our objects lack AGN features. Consequently, if
AGNs are present in the rest of our galaxies,  their luminosities should be much
fainter than 10$^{43}$ ergs s$^{-1}$ or alternatively, they must be heavily
obscured (contributing very weakly to the stellar continuum). The quality of our
data prevents us to provide a reliable pointlike+S\'ersic fit analysis of the
surface brightness distributions (as done by Daddi et al. 2005). However, an
argument  against the  biasing of our size estimates due to the AGN contribution
(if at all present)  is the fact that both at J and K the sizes of our objects
are very similar. The AGN should contribute much strongly in J--band (where
should be the emission lines H$_{\beta}$ and OIII at z$\sim$ 1.4) than in
K--band where there is not any significant line. The difference in sizes in J
and K in our objects are in agreement (within the error bars). For that reason,
we think that the AGN (if present) is not affecting (significantly) the size
estimates.

The galaxies analysed in this paper are already very massive at z$\sim$1.4 and
their stellar population properties are consistent with  passive fading.
However, there is no observational evidence for galaxies as massive and compact
as these in the local universe. Consquently, the high--z galaxies  have to 
increase their sizes since that redshift. This observational fact disagrees with
a scenario where the most massive and passive galaxies are fully assembled at
z$\sim$1.4 (i.e. a monolithic scenario). It is worth noting that, whatever 
channel is used for our galaxies to evolve in size, this process should take
place quickly (i.e. $\lesssim$ 2 Gyr), since galaxies with
M$_{\star}$$>$10$^{11}$h$_{70}^{-2}$M$_{\sun}$  at z$\sim$0.8  seem to be all
already in place (Cimatti et al. 2006), and to have sizes very similar to their
current values (McIntosh et el. 2005).

A very efficient size evolutionary mechanism (r$_e$$\propto$M$_{\star}^{1.3}$)
is found in dissipationless mergers with radial orbits (Boylan--Kolchin et al.
2006). In this process, galaxies do not evolve  parallel to the local relation
(r$_e$$\propto$M$_{\star}^{0.56}$). Interestingly, it turns out that these
radial disipationless mergers (along the filaments) of massive galaxies  is
thought to be the main channel of formation of the Brightest Cluster Galaxies
(BCG). Consequently, our very massive and extremely compact galaxies at
z$\sim$1.4 are very likely candidates to evolve into current BCGs. We have
explored whether the comoving number densities of the present day BCGs are in
agreement with the dry merger hypothesis. The Cole et al. (2001) local stellar
mass functions provides the following number densities for a Kroupa IMF:
$\sim$2.5x10$^{-5}$ Mpc$^{-3}$ for objects with
M$_{\star}$$>$3$\times$10$^{11}$M$_{\sun}$, and $\sim$4x10$^{-7}$ Mpc$^{-3}$ for
objects with M$_{\star}$$\sim$10$^{12}$M$_{\sun}$. If we assume that the number
densities of M$_{\star}$$>$3$\times$10$^{11}$M$_{\sun}$ objects at z$\sim$1.4 is
a 30\% than the present--day values, and consider that to reach the mass of a
BCG we need $\sim$4 of our galaxies, we would expect a comoving density of
$\sim$20x10$^{-7}$ Mpc$^{-3}$ for objects with
M$_{\star}$$\sim$10$^{12}$M$_{\sun}$ today. This is slightly higher than the
value measured for Cole et al. (2001) but works reasonable well due to the large
uncertainties. Consequently, we think a dry merger scenario can be considered as
a reasonable mechanism for the subsequent evolution of our galaxies (Khochfart
\& Burkert 2003; Dom\'{\i}nguez--Tenreiro et al. 2006). Alternative mechanisms
of galaxy evolution, like dissipative merging, will increase the mass of the
galaxies very effectively but will basically maintain unchanged the sizes (Dekel
\& Cox 2006). This will make the discrepancy in sizes between the high-z and the
local galaxies even larger. So, we think wet merging is disfavoured as an
evolutionary path for our objects.

An interesting open question is understanding how galaxies as massive as those
we are dealing with can be so compact in the past. Recently Khochfar \& Silk
(2005) have investigated the effect of  dissipation in major mergers within the
CDM paradigm. They find that  early-type galaxies at high redshifts merge from
progenitors that have  more cold gas available than their counterparts at lower
redshifts. As a  consequence, they claim that the remnant should be smaller in
size at high  redshift. Khochfar \& Silk (2006) have
predicted that the size of objects at z$\sim$1.5 with
M$_{\star}$$\gtrsim$5$\times$10$^{11}$h$_{70}$$^{-2}$M$_{\sun}$ is a factor of
$\sim$3 times smaller than local counterparts. These estimates agree very well
with our observations. If this scenario is correct, the progenitor galaxies that
merge to form massive spheroids galaxies are progressively  less and less
devoided of gas at lower redshift.

\vspace{-0.9cm}
\section{Acknowledgments}

We would like to thank S. Khochfar, C. Conselice, R.
Dom\'inguez-Tenreiro, J. O\~norbe and S. di Serego
for stimulating discussion. We thank the referee, E. Daddi, for a very useful 
report. G. F., Y. G., U. H., and R. B. acknowledge funding by the DFG (SFB 375).

\end{document}